\documentclass[prd,preprint,nopacs,nofootinbib,superscriptaddress]{revtex4}
\usepackage{amsfonts}
\usepackage{amsmath}
\usepackage{amsthm}
\usepackage{amssymb}
\usepackage{mathtools}
\usepackage{enumitem}
\usepackage{gensymb}
\usepackage{hyperref}
\usepackage{bm}
\usepackage{xcolor}
\usepackage{dsfont}
\usepackage{graphicx}
\usepackage{tipa}

\newcommand{\tetrad}{\text{\textschwa}}

\begin{document}
\title{Pregeometric First Order Yang-Mills Theory}
\author{Priidik Gallagher}
\email{priidik.gallagher@ut.ee}
\affiliation{Laboratory of Theoretical Physics, Institute of Physics, University of Tartu, W. Ostwaldi 1, 50411 Tartu, Estonia}

\author{Tomi Koivisto}
\email{tomi.koivisto@ut.ee}
\affiliation{Laboratory of Theoretical Physics, Institute of Physics, University of Tartu, W. Ostwaldi 1, 50411 Tartu, Estonia}
\affiliation{National Institute of Chemical Physics and Biophysics, R\"avala pst. 10, 10143 Tallinn, Estonia}
\affiliation{University of Helsinki and Helsinki Institute of Physics, P.O. Box 64, FI-00014, Helsinki, Finland}

\author{Luca Marzola}
\email{luca.marzola@cern.ch}
\affiliation{National Institute of Chemical Physics and Biophysics, R\"avala pst. 10, 10143 Tallinn, Estonia}

\begin{abstract}
    The standard description of particles and fundamental interactions is crucially based on a regular metric background. In the language of differential geometry, this dependence is encoded into the action via Hodge star dualization. As a result, the conventional forms of the scalar and Yang-Mills actions break down in a pregeometric regime where the metric is degenerate. This suggests the use of first order formalism, where the metric may emerge from more fundamental constituents and the theory can be consistently extended to the pregeometric phase. We systematically explore different realizations and interpretations of first order formalism, finding that a fundamental vector or spinor substructure brings about continuum magnetization and polarization as integration constants. This effect is analogous to the description of the cosmological dark sector in a recent self-dual formulation of gravity, and the similar form obtained for the first order Yang-Mills theory suggests new paths toward unification.
\end{abstract}

\maketitle

\section{Introduction}

The description of gravity in General Relativity (GR) is built on a 4-dimensional pseudo-Riemannian manifold supplying the fundamental field of interest: the metric. This describes distances and determines the curvature of spacetime through the Levi-Civita connection. The contemporary treatment of gauge fields can be taken to be just as geometric as that of gravity --- Yang-Mills theory is then built on top of the background manifold by considering the dynamical connections and curvature of a $G$-bundle pertaining to the chosen symmetry group $G$. Scalar and fermionic fields can also be described with a similar apparatus, which highlights a difference in the treatment of fermions and bosons. In fact, whereas the Dirac action is inherently of the first order, the actions of bosonic fields are usually given in the second order form and rely on the metric to operate the required contractions.

Thus, the formulation of Quantum Field Theory (QFT), and the Standard Model of particle physics in particular, presupposes a (constant, Minkowski) metric background~\cite{Weinberg:1996}, and in the context of GR this background is promoted to a dynamical field. Consequently, both frameworks completely break down in the hypothetical situation where the metric field becomes degenerate. This possibility was already considered by Einstein and Rosen, in their attempt of providing a geometrical description of elementary particles through ``bridges'' characterised by a vanishing metric determinant, $g=0$~\cite{Einstein:1935tc}. Later, such solutions and their topology have played an important part in attempts at quantum geometrodynamics~\cite{Wheeler:1957mu,Henneaux:1981su}. In particular, it has been proposed that in quantum gravity, the {\it ground state of the metric} field should be  $g_{\mu\nu}=0$~\cite{Witten:1988hc,Horowitz_1991}. We point out a semantic inconsistency which occurs if the definition of a metric is taken to require an \textit{invertible}, symmetric and covariant tensor. In the case that one insists upon a metric theory in this sense, the physical implication of an ``ametric'' phase, i.e. the loss of causal structure, then manifests in the inevitable non-locality of the ultraviolet completion, e.g. in terms of infinite-derivative~\cite{Biswas:2011ar} or fake degrees of freedom~\cite{Anselmi:2018kgz}.

A vanishing ground state for the metric was explicitly realised in a recent {\it pregeometric} gravity theory~\cite{Zlosnik_et_al:2018,Koivisto:2019ejt}. The related pregeometry programme proposes that the metric, or the (co)frame, is emergent and composed of other, potentially purely fermionic fields. The framework is in line with earlier studies such as that of Ref.~\cite{Akama:1978}, later invoked within unification~\cite{Amati_Veneziano:1981}, spinor gravity~\cite{Hebecker_Wetterich:2003,Wetterich:2021},
analogue gravity~\cite{VOLOVIK1990222}, time-space asymmetry~\cite{Wetterich:2021b}, lattice regularization~\cite{Diakonov:2011} and cosmology~\cite{Akama:1981kq,Wetterich:2021cyp}. By introducing the emergent coframe via the exterior covariant derivative of a Lorentz vector (or, a bispinor) in a Cartan-geometric language,  Ref.~\cite{Zlosnik_et_al:2018} found that  the consistent General-Relativistic solutions are immediately accompanied by dust interpretable as dark matter\footnote{In a unimodular version of this theory, both the cold dark matter (CDM) and the cosmological constant $\Lambda$ arise as integration constants~\cite{Gallagher:2021tgx}. This provides, to our knowledge, the unique candidate for a $\Lambda$CDM {\bf theory} of cosmology. The topological origin of the CDM, due to the existence of a $g_{\mu\nu}=0$ phase, had been anticipated in the work of Ba{\~n}ados~\cite{Banados:2007qz}.}. In the ground state, the gauge connection is arbitrary. To support a Minkowski background, the Lorentz vector spontaneously breaks the symmetry of the theory by acquiring a time-like expectation value and the gauge connection configuration has both torsion and curvature. Thus, in terms of the more primitive fields, there is a non-trivial structure underlying the Minkowski metric background.

As we show in the present paper, the fact that the metric could be emergent and admit a singular phase forces to question the traditional descriptions of elementary fields. In fact, although the fermionic sector poses no problem, in a singular metric phase the conventional actions for scalar and Yang-Mills theory have to be abandoned due to the presence of a potentially singular inverse metric in the Hodge dualization. To overcome the problem, we systematically study the possibilities offered by the first order formalism, seeking forms for the bosonic actions which recover the usual equation of motions and are suitable for a possible pregeometric regime.

Several approaches have been investigated before. For instance, Ref.~\cite{Pagels:1984}, working in Euclidean signature, constructed matter actions invariant under $O(5)$, independent of the metric and connection via introducing auxiliary fields, while Ref.~\cite{Wilczek:1998} used a ``preferred volume'' formalism, without requiring general covariance, but only covariance under volume-preserving reparametrizations, likewise introducing an auxiliary field, which coincides with the inverse vierbein after symmetry breaking. Ref.~\cite{Westman_Zlosnik:2013} studied a Yang-Mills-Cartan action, where the gauge field included a Cartan index, and was the initial basis for developing the approach here. The Yang-Mills-Cartan action is included in their Cartan-unified theory, but separate de Sitter gauge invariance requires assuming that the contact vector is constant. Thus, the previous approaches have been based on  5-dimensional extensions of the 4-dimensional orthogonal symmetry. Starting from Lorentz symmetry, this article will instead discuss actions that essentially are alternative realizations of a first order Yang-Mills theory, that now with polynomially simple actions appear to be particularly applicable in the study of gravity. A similar approach to scalar field theory is possible as well, and is only briefly mentioned, but it deserves more future analysis.

For terminological clarity, it may be useful to distinguish our approach from the {\it premetric} discourse in the Literature. The basic idea is the same: removing the metric from the fundamental equations of physics. The premetric program, put forward by Kottler in 1922, has developed into an axiomatic framework for analyzing and constructing the structure of a theory, beginning from the identification of the conserved quantities, and avoiding the reference to any metrical concepts as far as possible~\cite{Hehl:2016glb}. In principle, this framework allows to consider very general theories which do not even necessarily admit a Lagrangian formulation. In practice, however, the metric is often introduced at the stage when the theory is made predictive by postulating its constitutive law. Premetric electrodynamics has been very well studied, a classic textbook reference is~\cite{Hehl_Obukhov:2003}, and a shorter overview is~\cite{Hehl_Itin_Obukhov:2005}, for extension to gravity see~\cite{Obukhov:2019, Koivisto_Hohmann_Marzola:2021}. Yet, an
extension to Yang-Mills theory appears to be missing.

The pregeometric theory that we pursue is, more particularly, a theory that is based on an invariant, polynomial action principle which remains well-defined when the inverse metric (that only emerges, potentially highly non-polynomially, as a composite of more primitive fields) does not exist, becomes degenerate or even singular. Thus, we could state that what we mean by a pregeometric theory is a premetric theory satisfying the two key axioms: 1) formulation as an action principle and 2) viability of the ``ametric'' phase. We shall make connection with the existing constructions of premetric electromagnetism, wherein the electromagnetic excitation is introduced in conjunction with the axiom of charge conservation and the form of the excitation is finally postulated (with or without invoking a metric) through the constitutive law, by instead promoting the excitation to a dynamical field. Thus, the constitutive law is the consequence of dynamics and only valid on-shell, and moreover, the usual relation between the excitation and the field strength may only emerge in the metric phase. We then proceed to explore the possibility of reducing, together with the metric field, the new dynamical excitation field into more primitive substructure.

The article is structured as follows. In Section~\ref{sec:preliminaries} we will briefly go over our conventions and set the formalism used to describe matter fields. The bulk of the article, in section~\ref{sec:Yang_Mills}, is devoted to introducing the first order Yang-Mills actions and studying them. In particular, several interpretational questions will be looked over, and the vector substructure implications are investigated, producing analogous results to~\cite{Zlosnik_et_al:2018}. Section~\ref{sec:unification_questions} brings attention to the similarity of this first-order Yang-Mills theory and self-dual Palatini gravity and goes over some questions involving unification of gravity in first order formalism, before reaching the conclusion.

\section{\label{sec:preliminaries}Classical Theory of Matter and Gravity}
\subsection{Some Mathematical Preliminaries}
Our conventions are $(\eta_{ab})=\mathrm{diag}(-,+,+,+)$ and $\epsilon_{0123}=-\epsilon^{0123}=-1$. Lorentz indices are in Latin, with the spatial components capitalized, while Greek indices refer to the coordinate basis, or arbitrary basis in this section. We implicitly use natural units $c=\hbar=1$, generally barring numerical coefficients if they do not modify the analysis nor the dynamics; coefficients and coupling constants are introduced in section~\ref{sec:unification_questions} for comparing GR with Yang-Mills theory.

A few concepts that should be emphasized are more clear in arbitrary $n$ dimensions, but will be restricted to 4 dimensions later on. Then, a general $p$-form
\begin{equation}
	\omega=\frac{1}{p!}\omega_{\mu_1\ldots\mu_p}\vartheta^{\mu_1}\wedge\ldots\wedge\vartheta^{\mu_p}.
\end{equation}
It is important to note that the Levi-Civita symbols $\epsilon^{\mu_1\ldots\mu_n}$ and $\epsilon_{\mu_1\ldots\mu_n}$ themselves are a premetric concept, arising from differential forms of maximal rank
\begin{equation}
	\vartheta^{\mu_1}\wedge\ldots\wedge\vartheta^{\mu_n}=\epsilon^{\mu_1\ldots\mu_n}\vartheta^{0}\wedge\ldots\wedge\vartheta^{n-1}=\epsilon^{\mu_1\ldots\mu_n}\hat{\epsilon}
\end{equation}
and their maximal interior product
\begin{equation}
	\epsilon_{\mu_1\ldots\mu_n}=\mathrm{sgn}(g)\ \tetrad_{\mu_n}\lrcorner \ldots\lrcorner \tetrad_{\mu_1} \hat{\epsilon},
\end{equation}
where the 1-forms $\vartheta^{\mu_i}$ are an arbitrary cobasis with $\tetrad_{\mu_i}$ its respective vector basis; note the addition of the sign is here only conventional, not strictly required for constructing the symbol. Without a metric, there is no immediate correspondence between the two symbols, which is sometimes notationally emphasized. The placement (or omission) of the sign of the metric determinant is the primary point of contention between various Levi-Civita symbol and tensor conventions. Here it is added to the symbol $\epsilon_{\mu_1\ldots\mu_n}$, but alternatively it could instead be added to the symbol $\epsilon^{\mu_1\ldots\mu_n}$ or either of the Levi-Civita tensors. In practical terms, mainly the Levi-Civita tensors are used, but as can shortly be seen, in an orthonormal frame they coincide with the Levi-Civita symbols up to sign conventions. Furthermore, note that in the paper we further assume Lorentz symmetry as a starting point, therefore the symbols $\epsilon_{abcd}$ and $\eta_{ab}$ are available as invariants of the symmetry, and in particular, $\eta_{ab}$ can be seen not as a field on spacetime, but simply as an invariant of Lorentz symmetry.

Furthermore, the chain of interior products $\hat{\epsilon}_{\mu_1}=\tetrad_{\mu_1}\lrcorner\hat{\epsilon},\hat{\epsilon}_{\mu_1\mu_2}=\tetrad_{\mu_2}\lrcorner \tetrad_{\mu_1}\lrcorner\hat{\epsilon},\ldots$ yields a differential form basis equivalent to $\vartheta^{\mu_1},\vartheta^{\mu_1}\wedge\vartheta^{\mu_2},\ldots$; see e.g.~\cite{Hehl_Obukhov:2003} for more discussion. Therefore it is also possible to expand differential forms as
\begin{equation}
	\omega=\frac{1}{(n-p)!}\omega^{\mu_1\ldots\mu_{n-p}}\hat{\epsilon}_{\mu_1\ldots\mu_{n-p}}.
\end{equation}

Moving between a differential form description and the usual index formalism can be realized with the $\diamond$-dual density, which establishes a correspondence between $p$-forms and totally antisymmetric tensor densities of weight $+1$ and type $(n-p,0)$, i.e. $(n-p)$ vectors. Generally in terms of basis vectors in $n$ dimensions
\begin{equation}
	\diamond(\hat{\epsilon}_{\mu_1\ldots\mu_p})=\delta^{\nu_1\ldots\nu_p}_{\mu_1\ldots\mu_p}\tetrad_{\nu_1}\otimes\ldots\otimes \tetrad_{\nu_p},
\end{equation}
thus for a general $p$-form $\omega$ the $\diamond$-dual tensor density
\begin{equation}\label{eq:diamond_dual}
	\diamond\omega=\frac{1}{(n-p)!}\omega^{\mu_1\ldots\mu_{n-p}}\tetrad_{\mu_1}\otimes\ldots\otimes \tetrad_{\mu_{n-p}},
\end{equation}
with the components
\begin{equation}
	\omega^{\mu_1\ldots\mu_{n-p}}=\frac{1}{p!}\omega_{\mu_{n-p+1}\ldots\mu_n}\epsilon^{\mu_1\ldots\mu_n}.
\end{equation}
Note this duality between differential forms and tensor densities does not yet use the metric, as neither the expansion with respect to the Levi-Civita dual basis $\hat{\epsilon}_{\mu_1\mu_2\ldots}$ nor the Levi-Civita symbols involve the metric, and is therefore viable in a premetric description.

However, establishing the Hodge $*$-duality between differential forms of rank $p$ and $n-p$ does require the metric, see in components
\begin{equation}\label{eq:Hodge_dual}
	*\omega=\frac{\sqrt{-g}}{p!(n-p)!}\omega^{\mu_1\ldots\mu_p}\epsilon_{\mu_1\ldots\mu_p\mu_{p+1}\ldots\mu_n}\vartheta^{\mu_{p+1}}\wedge\ldots\wedge\vartheta^{\mu_n},
\end{equation}
where the metric determinant appears explicitly and the inverse metric was used to raise indices. Likewise, the metric appears in the Levi-Civita tensors 
\begin{align}
	\varepsilon_{\mu_1\ldots\mu_n}&=\sqrt{\vert g\vert}\epsilon_{\mu_1\ldots\mu_n},\\
	\varepsilon^{\mu_1\ldots\mu_n}&=\frac{1}{\sqrt{\vert g\vert}}\epsilon^{\mu_1\ldots\mu_n}.
\end{align}
However, note the volume form in 4 dimensions
\begin{equation}
	\mathrm{Vol}=\frac{1}{4!}\varepsilon_{\mu\nu\rho\sigma}\mathrm{d}x^\mu\wedge\mathrm{d}x^\nu\wedge\mathrm{d}x^\rho\wedge\mathrm{d}x^\sigma=\frac{1}{4!}\epsilon_{abcd}e^a\wedge e^b\wedge e^c\wedge e^d,
\end{equation}
where in an orthonormal cobasis it loses explicit reference to the metric, as the determinant $\eta=-1$.

\subsection{Matter Actions in Differential Forms}
Equipped with the apparatus of differential geometry we review the standard actions of matter and gravity in the framework of Lorentz symmetry, paying special attention to the role of the metric. For a more thorough discussion of possible geometric descriptions of the background and their possible equivalence, we refer the reader to Ref.~\cite{Jimenez:2019woj}.

The simplest case of a massless Dirac fermion can be described in Riemann-Cartan geometry through the Dirac spinor action
\begin{equation}
	S_\psi=-\int\bar{\psi}\gamma\wedge*i\mathrm{D}\psi=\int\mathrm{d}^4{x}\sqrt{-g}\ \bar{\psi}\gamma^\mu i\mathrm{D}_\mu\psi,
\end{equation}
where $\gamma=\gamma_a e^a$ is a 1-form and $\gamma_a$ are the Dirac gamma-matrices, while the exterior covariant derivative (barring possible gauge interactions) acts on the spinor 0-form as
\begin{equation}
	\mathrm{D}\psi=\mathrm{d}\psi-\frac{i}{2}\omega_{ab}\bigg(-\frac{i}{4}[\gamma^a,\gamma^b]\bigg)\psi.
\end{equation}
The action can be straightforwardly extended to include a Dirac mass term  $\bar{\psi}m\psi\mathrm{Vol}$, possibly generated after the spontaneus breaking of gauge symmetries via the Higgs mechanism. The explicit mathematical construction of spinor theory on curved spacetime is lengthier. A simple exposition is that the 0-forms are spinor-valued in Minkowski space tangent to the background manifold\footnote{We refer the reader to e.g. Ref.~\cite{Lawson_Michelsohn:1989} for a discussion pertaining to spin geometry and the construction of spinor bundles. Notably, it is not possible to define such structure on completely arbitrary manifolds, as generally there can be topological obstructions. Pregeometry could liberate from such obstructions. (One can comb a hairy ball if it is allowed to have a bald spot.)}.

Scalar field and Yang-Mills theories usually rely on second order actions, the kinetic terms being, respectively,
\begin{align}
	S_\phi&=-\frac{1}{2}\int\mathrm{d}\phi\wedge*\mathrm{d}\phi,\label{eq:second_order_scalar}\\
	S_A&=-\frac{1}{2}\int\mathrm{Tr}\big(F\wedge*F\big).\label{eq:second_order_YM}
\end{align}
Scalar fields $\phi$ are just functions, i.e. 0-forms on a manifold, while Yang-Mills fields $A$ are the $G$-connection 1-forms of the corresponding gauge symmetry group $G$ with Lie algebra $\mathfrak{g}$, and the respective curvature 2-form is given as
\begin{equation}
	F=\mathrm{d}A+A\wedge A.
\end{equation}
There are several ways to describe connections, for instance through a $G$-invariant horizontal distribution, as a family of 1-forms etc. Here, implicitly working with $G$-bundles $E$ over $M$, and purely for a general description, we can assume that $A$ is a section of $\mathrm{End}(E)$ valued in $\mathfrak{g}$. Therefore, locally $A$ is a Lie algebra valued 1-form on the $G$-bundle. Most of the various descriptions are equivalent, emphasizing different properties, and it is possible to consistently work in local charts if necessary. Proceeding forward, variation with respect to $\phi$ produces the Klein-Gordon equation, while the variation with respect to $A$ yields the inhomogeneous Yang-Mills equation $\mathrm{D}*F=J$. The homogeneous equation is trivial as the Bianchi identity $\mathrm{D}F=0$ is satisfied by construction of the curvature 2-form.

Finally, the Palatini gravity action is
\begin{equation}
	S_{e^a,\omega_{ab}}=\frac{1}{2\kappa}\int e^a\wedge e^b\wedge *R_{ab},
\end{equation}
where $R_{ab}$ is the curvature 2-form of the spin connection 1-form $\omega_{ab}$. The basic variables are the coframe $e^a$ and the spin connection $\omega_{ab}$, variation with respect to the first producing the Einstein equations, while the equations of motion of the latter fix the connection used in calculating curvature and used in the energy-momentum. In particular, when there is no contribution to spin density, the torsion vanishes and the connection reduces to Levi-Civita.

As we can see, Hodge dualization appears in all the actions reported above, thereby apparently preventing us from immediately using them in a regime of the theory where the inverse metric, used in the dualization, is not available. Regarding this, we remark that the problem can actually be circumvented in pure spinor theories, as the action can be spelled out explicitly in an arbitrary orthonormal coframe
\begin{equation}
	S_\psi=\int\epsilon_{abcd}e^a\wedge e^b\wedge e^c\wedge \bar{\psi}\gamma^d i\mathrm{D}\psi,
\end{equation}
without using the inverse metric nor the tetrad. On top of that, the Hodge star operator in the gravitational action is particularly amiable in self-dual formulation, discussed in section~\ref{sec:unification_questions}. The relevant formulation of scalar and Yang-Mills theory, suitable for a pregeometric regime independent of the metric, is not as well-known but can be resolved in a first order theory, as we will show below.

\section{\label{sec:Yang_Mills}The Yang-Mills Kinetic Cycle}
\subsection{An auxiliary 1-form}
To describe the dynamics of the $G$-connection gauge field $A$ for a (generally non-Abelian) gauge group $G$, consider the action, for brevity neglecting overall coefficients,
\begin{equation}\label{eq:ua_action}
	S=\int\mathrm{Tr}_G\bigg(\frac{1}{2}\epsilon_{abcd}u^a\wedge e^b\wedge u^c\wedge e^d + \eta_{ab}u^a\wedge e^b\wedge F\bigg),
\end{equation}
where $u^a$ is an auxiliary Lie algebra $\mathfrak{g}$ valued 1-form, transforming as
\begin{align}
	A&\to g A g^{-1} + g\mathrm{d}g^{-1}\Rightarrow F\to g F g^{-1},\\
	u^a&\to \mathrm{Ad}_g u^a=g u^a g^{-1}.
\end{align}
The action is thus by construction invariant under local gauge transformations, local Lorentz transformations and diffeomorphisms. Likewise, it is polynomial and only involves derivatives up to first order. Importantly, the action only makes use of the fundamental objects available: a coframe\footnote{Note in particular that the (prototype) coframe $e^a$ used here need not be regular everywhere and is only required to allow a proper expansion with respect to it in the nonsingular phase. In this sense, the coframe 1-forms produce the geometric structure of interest, rather than just being part of the description of spacetime. For conciseness we will refer to $e^a$ just as a coframe rather than a pseudo-coframe.} $e^a$, a 1-form $u^a$, the vector potential $A$ (contained in its field strength 2-form $F$), and the invariants $\epsilon_{abcd}$ and $\eta_{ab}$ of the Lorentz group. A second coframe is present in the conformal extension of the Lorentz symmetry~\cite{Koivisto:2019ejt}, see also e.g.~\cite{Muhwezi:2020vpw,Zlosnik:2016fit}. However, here we simply consider the (presumably broken) symmetry $SO(1,3)\times G$, and shall return to discuss paths to unification
in section \ref{sec:unification_questions}.

The corresponding equations of motion of $A$ and $u^a$ are respectively
\begin{align}
    \mathrm{D}(\eta_{ab} u^a\wedge e^b) &= J,\\
	\epsilon_{abcd}u^b\wedge e^c\wedge e^d + e_a\wedge F &= 0.\label{eq:EoM_ua}
\end{align}
Here $J$ is the current 3-form, which is generally sourced by the matter terms in the total action\footnote{When a metric is available, $J$ is usually taken to be the Hodge dual of a current 1-form, so $J=*j$.}. The first equation is just the prototype inhomogeneous Yang-Mills equation, while the second equation, as we will show, enforces that on-shell $\eta_{ab} u^a\wedge e^b$ be related to the Hodge dual of gauge curvature.

The classical equivalence of this formulation with the usual Yang-Mills theory can be shown by considering $\diamond$-dual densities of the differential forms, as defined in Eq.~\eqref{eq:diamond_dual}. In our case, the analysis will be done in a prototype orthonormal coframe $e^a$, which is required to be a proper coframe in the non-singular phase $\mathrm{det}(\eta_{ab}e^a{}_\mu e^b{}_\nu)\neq0$. The differential forms $u^a$ and $F$ can thus be expanded in the basis provided by $e^a$ and we can employ the Minkowski metric to raise and lower indices as necessary. In this case, in components
\begin{equation}
	\epsilon_{abcd}u^b{}_i\epsilon^{icdk}+\frac{1}{2}\eta_{ab}F_{cd}\epsilon^{bcdk}=0,
\end{equation}
which is
\begin{equation}
	2u^b{}_i \delta^{ik}_{ab}=2(u^k{}_a - \delta_a^k u^i{}_i)=\frac{1}{2}\eta_{ab}F_{cd}\epsilon^{bcdk}.
\end{equation}
Tracing over the indices $a$ and $k$ implies $u^i{}_i=0$, and finally utilizing the Minkowksi metric to lower indices implies $u_{(ab)}=0$. The remaining antisymmetric part, $u_{ab}=u_{[ab]}$, can be reorganized to
\begin{equation}\label{eq:ua_Hodge}
	u_{ak}=-\frac{1}{4}F^{ij}\epsilon_{ijak},
\end{equation}
which coincides with the Hodge star in an orthonormal coframe. As $u^a=u^a{}_k e^k$, in the nonsingular phase
\begin{equation}\label{eq:F_ua_dual}
	\eta_{ab}u^a\wedge e^b=*F.
\end{equation}
Substituting the above relation back into the prototype inhomogeneous equation produces the usual Yang-Mills equations. As $\mathrm{D}F=0$ is trivially satisfied, this first order formulation is classically equivalent to usual Yang-Mills theory by realization of a two-step ``kinetic cycle''. Likewise, the on-shell action neatly coincides with the usual Yang-Mills action, as multiplying Eq.~\eqref{eq:EoM_ua} by $u^a$ and taking the trace immediately yields
\begin{equation}
	\mathrm{Tr}\Big(\epsilon_{abcd}u^a\wedge e^b\wedge u^c\wedge e^d\Big)=\mathrm{Tr}\Big(-\eta_{ab}u^a\wedge e^b\wedge F\Big),
\end{equation}
and the results follows through Eq.~\eqref{eq:F_ua_dual}.

The action in Eq.~\eqref{eq:ua_action} can be considered to be pregeometric in the sense that it remains well defined, as do the corresponding equations of motion, even if the physical metric $g_{\mu\nu}=e^a{}_\mu e^b{}_\nu\eta_{ab}$ is singular. In fact, neither the action nor the equations of motion depend on the inverse metric
\begin{equation}
	g^{\mu\nu}=\frac{4\epsilon^{\alpha_1\alpha_2\alpha_3\mu}\epsilon^{\beta_1\beta_2\beta_3\nu}g_{\alpha_1\beta_1}g_{\alpha_2\beta_2}g_{\alpha_3\beta_3}}{\epsilon^{\alpha_1\alpha_2\alpha_3\alpha_4}\epsilon^{\beta_1\beta_2\beta_3\beta_4}g_{\alpha_1\beta_1}g_{\alpha_2\beta_2}g_{\alpha_3\beta_3}g_{\alpha_4\beta_4}}.
\end{equation}
The action is not completely premetric, however, as it still requires the Minkowski metric $\eta_{ab}$. The general similarity of our approach to topological QFT, topological Yang-Mills theory, and BF theory has to be noted, for further inspiration.

Heuristically, to emulate the Hodge dualization without having a regular metric at hand, it makes sense to begin with an expansion in the prototype orthonormal coframe, allowing for a possible singular phase. In order to recover the ordinary Yang-Mills theory, first setting aside possible topological terms, any prototype kinetic term $X\wedge F$ will require $X$ to become the dual field strength 2-form $*F$, which can easiest be done by introducing explicit Lorentz indices and 1-form substructure to $X$, the simplest substitution being $X\to \eta_{ab} u^a\wedge e^b$, as in Eq.~\eqref{eq:ua_action}. This can then be coupled with the Levi-Civita dualized basis as $\epsilon_{abcd}u^a\wedge e^b\wedge u^c\wedge e^d$, similarly to how the Hodge dualization~\eqref{eq:Hodge_dual} connects differential form components to a dual basis. The procedure could be extended by introducing more auxiliary fields or Lagrange multipliers, thereby lengthening the cycle, but it does not appear helpful at present stage. Likewise, it is by no means the only method to construct a pregeometric Yang-Mills theory, as discussed earlier.

Other approaches can be thought of too. For instance, introducing complex structure and instead considering the field strength $F$ as a fundamental field, the Yang-Mills equations in 4 dimensions are equivalent to the system
\begin{equation}\label{eq:dual_equations}
	\mathrm{D}{}^\pm F=\mp\frac{i}{2}J,
\end{equation}
as $**F=-F$ in Minkowski signature. In particular, when considering $U(1)$ theory, the covariant derivative is replaced by the exterior derivative, which then immediately implies the inhomogeneous Maxwell equation and that $F$ is closed, therefore under suitable topological assumptions by de Rham's first theorem (cf. the Poincare lemma), $F$ is also exact. In the non-Abelian case \eqref{eq:dual_equations} does not work as neatly, but more crucially, this approach implicitly still requires the use of the Hodge dual in the definition of ${}^\pm F$. It is not clear whether it is possible to work around this.

Initially, the proposed ``kinetic cycle''  was developed purely from the ideas in Ref.~\cite{Westman_Zlosnik:2013}, but without introducing any Cartan ``rolling'' indices; note that the action~\eqref{eq:ua_action} does not introduce more degrees of freedom than employing rolling indices would do. However, the action~\eqref{eq:ua_action} essentially is a realization of first order Yang-Mills theory, applicable in arbitrary spacetime and in presence of metric singularities. The utility of first order (Abelian) Yang-Mills theory in degenerate spacetimes was already analyzed in Ref.~\cite{Kaul:2019} and earlier in context of degenerate tetrads discussed in Ref.~\cite{Canarutto:1998}, despite without the explicit continuation to non-Abelian theory. More expansive study of first order theory itself goes back to at least the 1970-s~\cite{Halpern:1977a,Halpern:1977b} in relation to strong coupling effects, but it has also appeared even longer ago, e.g.~\cite{Kibble:1961} in an appendix or see (the republication)~\cite{Arnowitt:1962hi} in remark and comparison of Maxwell electromagnetism with gravity, and is more generally related to the Ostrogradski procedure of lowering derivative order~\cite{Deser:2006, Kiriushcheva_Kuzmin:2007}. It has also been employed in relation to the Duffin-Kemmer formulation~\cite{Okubo_Tosa:1979, Okubo_Tosa:1981}, or computation of loop effects~\cite{McKeon:1994} in emphasis of the simpler vertex structure. Proving the classical equivalence of first order and ordinary formulation of Yang-Mills theories is rather straightforward. Equivalence at the quantum level was recently studied via vacuum functionals~\cite{Lavrov:2021}, and earlier discussion can be found in e.g.~\cite{Kiriushcheva_Kuzmin_McKeon:2012}, while other recent results include study into renormalization~\cite{Brandt_Frenkel_McKeon:2018} and consistency conditions related to Green's functions and ultraviolet divergences~\cite{McKeon_et_al:2020}. Furthermore, as an aside to pregeometric deliberation, first order theory has been formulated as a deformation of topological BF theory~\cite{Cattaneo_et_al:1998}.

In the electromagnetic $U(1)$ theory, the 1-form $u^a$ can also be interpreted in terms of the electromagnetic excitation $H$, appearing from electric current conservation as $\mathrm{d}J=0\Rightarrow J=\mathrm{d}H$. The excitation, both in the equations of motion and the action, appears on a premetric level, see Ref.~\cite{Hehl_Obukhov:2003} for details. In this light, rather than axiomatically defining correspondence between the excitation and dual field strength via the electromagnetic spacetime relation, in a first order theory this correspondence appears because of the specific form of the action and the excitation itself can be regarded as a fundamental field of the theory.

In global formulations, $u^a$ is immediately reminiscent of an extra coframe field, albeit Yang-Mills charged. This can be further related to bimetric theory, see Ref.~\cite{Pavlovsky:2002} for proposals, although for $u^a$ to be considered a proper coframe-like object, the implications of expansion w.r.t. $u^a$ require investigation\footnote{Interestingly, the action features precisely the partially massless interaction term of the Hassan-Rosen ghost-free bimetric theory~\cite{Hassan:2012gz}. Besides the partially massless term, there exist two other viable interactions~\cite{Hassan:2011zd}. It might be interesting to explore how including these terms would modify the first order Yang-Mills theory, and whether the bimetric modified gravity could perhaps be interpreted in this connection with particle physics.}. 
Curiously enough, the 1-form $u^a$ allows to define a Yang-Mills derived pseudometric for any Yang-Mills theory
\begin{equation}
	g_{\mu\nu}^\text{YM}=\mathrm{Tr}(u^a{}_\mu u^b{}_\nu\eta_{ab})\,.
\end{equation}
This is not necessarily canonical, however, as it is possible to derive similar structure from the interior product of the field strength 2-form $F$ with respect to a vector basis, and likewise the interpretation of $g_{\mu\nu}^\text{YM}$ is unclear. Quite interestingly, in $D=3$ dimensions similar metric construction connects with gravity rather closely, see~\cite{Reinhardt:1996,Lunev:1996}.

Finally, as expected, in a gravitational context the energy-momentum derived from $e^a$ also agrees with the usual Yang-Mills energy-momentum tensor. Variation w.r.t. the coframe $e^a$, yields the canonical energy-momentum 3-form
\begin{equation}
	\theta_a:=\mathrm{Tr}_G\Big(\epsilon_{abcd}e^b\wedge u^c\wedge u^d - u_a\wedge F\Big).
\end{equation}
One can also derive the energy-momentum from \eqref{eq:second_order_scalar} and \eqref{eq:second_order_YM}, and the equivalence is simplest seen via index calculations in the dual densities, investigating the component expression. Assuming $u^a$ is on-shell, so that Eq.~\eqref{eq:ua_Hodge} holds, and a regular metric phase, algebraic manipulation yields altogether
\begin{equation}
	\mathrm{Tr}\Big(\varepsilon_{abcd}u^c{}_i u^d{}_j \varepsilon^{bijk} -  u_a{}_b\frac{1}{2}F_{ij}\varepsilon^{bijk}\Big)=
	\mathrm{Tr}\Big(-(F_{ai}F^{ki}-\frac{1}{4}\delta_a^k F^{ij} F_{ij})\Big).
\end{equation}
That is
\begin{equation}
	T^{\mu\nu}\sim \mathrm{Tr}\Big(F^{\mu\rho}F^\nu{}_\rho-\frac{1}{4}g^{\mu\nu}F_{\rho\sigma}F^{\rho\sigma}\Big),
\end{equation}
therefore, up to conventions, the energy-momentum tensor agrees with that of the usual theory.

\subsection{A Yang-Mills charged transformation}
Equivalently, it is possible to instead introduce a Lie algebra valued 0-form $G^{ab}$ with antisymmetric Lorentz indices, via the action
\begin{equation}\label{eq:Gab_action}
	S=\int\mathrm{Tr}\Big(\frac{1}{24}G^{ab}G^{cd}\eta_{ac}\eta_{bd}\epsilon_{ijkl}e^i\wedge e^j\wedge e^k\wedge e^l + G^{ab}\eta_{ac}\eta_{bd}e^c\wedge e^d\wedge F\Big).
\end{equation}
The resulting equations of motion w.r.t. $A$ and $G^{ab}$ are respectively
\begin{align}
	\mathrm{D}(G_{ab}e^a\wedge e^b)&=J,\\
	\frac{1}{12} G^{ab}\varepsilon_{ijkl}e^i\wedge e^j\wedge e^k\wedge e^l+e^a\wedge e^b\wedge F&=0.
\end{align}
The analysis mirrors that of the previous section, going to the dual basis yields
\begin{equation}
	G_{ab}e^a\wedge e^b=*F,
\end{equation}
and e.g. analyzing the energy-momentum would proceed in a similar way. Likewise the action \eqref{eq:Gab_action} makes no reference to the inverse metric, thus allows for a singular phase and could be considered pregeometric.

The main difference lies in the interpretation, discussed in section \ref{sec:unification_questions}. Equivalently we can construct
\begin{equation}
	S=\int\mathrm{Tr}\Big(\frac{1}{4}(G^{ab}\eta_{ac}\eta_{bd}e^a\wedge e^b)\wedge (G^{ij}\epsilon_{ijkl} e^k\wedge e^l) + (G^{ab}\eta_{ac}\eta_{bd}e^c\wedge e^d)\wedge F\Big),
\end{equation}
where the coupling of $G^{ab}$ to the surface element $e^a\wedge e^b$ is more explicit, similar to the appearance of $u^a\wedge e^b$ in the 1-form approach. Either approach is classically equivalent to second order Yang-Mills theory, and there is evidence of quantum equivalence as well. Particularly in flat space, the differential form description can be reduced to (a variant of) first order Yang-Mills theory in the usual index formalism, for which the study of quantum properties and equivalence was discussed earlier, see e.g.~\cite{Lavrov:2021}. The quantum properties of the first order formalism while remaining in curved spacetime require further investigation, however, and a deeper overview of applying the many possible quantization schemes promises to be insightful as well.

\subsection{Vector substructure}
In analogy with Ref.~\cite{Zlosnik_et_al:2018}, we can produce the coframe from a single Yang-Mills charged vector $\phi^a$, such that
\begin{equation}
	u^a=\mathrm{D}\phi^a=\mathrm{d}\phi^a+\omega^a{}_b\phi^b+[A,\phi^a].
\end{equation}
Therefore, rather than introducing a Lie-algebra valued 1-form $u^a$, only a single Lorentz vector $\phi^a$ is postulated. Explicitly, the action becomes
\begin{equation}\label{eq:phia_action}
	S=\int\mathrm{Tr}_G\bigg(\frac{1}{2}\epsilon_{abcd}\mathrm{D}\phi^a\wedge e^b\wedge \mathrm{D}\phi^c\wedge e^d + \eta_{ab}\mathrm{D}\phi^a\wedge e^b\wedge F\bigg),
\end{equation}
with the equations of motion for $A$ and $\phi^a$ respectively
\begin{align}
	\varepsilon_{abcd}[\phi^a,\mathrm{D}\phi^b] \wedge e^c\wedge e^d
	+ e_a\wedge [\phi^a,F]
	+ \mathrm{D}(\mathrm{D}\phi^a\wedge e_a)&=J,\\
	\mathrm{D}\Big(\varepsilon_{abcd}\mathrm{D}\phi^b\wedge e^c\wedge e^d+e_a\wedge F\Big)&=0.
\end{align}
Further, note that it is possible to consider a single Dirac spinor $\psi$ instead of a vector $\phi^a$, similarly to how various spinor-pregeometric approaches work with coframes. This is most clear in the commutative case of $U(1)$ theory, where we could consider a substitution of the type $\phi^a\to\bar{\psi}\gamma^a\psi$, with the objects in the adjoint representation being invariant under transformations. The non-Abelian case, however, requires more structure than a single spinor $\psi$, cf. extra 
Yang-Mills indices\footnote{An attractive possibility would be to consider $\phi$ as a $G$-vector, and $\bar{\psi}$ as its dual $G$-vector, so that $\bar{\psi}\gamma^a\psi$ would have its indices in the adjoint as desired. This would essentially realise the same trick internally as we are now performing externally, by considering the Lorentz vector $\phi^a$ instead of the Lorentz adjoint $G^a{}_b$. The trick would considerably reduce the number of independent variational degrees of freedom. However, it remains to be investigated whether the gauge-invariant degrees of freedom in the (dual) field strength can be consistently encoded within one $G$-vector spinor (or whether we may would have to e.g. consider the $\psi$ and $\bar{\chi} \neq \bar{\psi}$ as independent).}.  

Although the analysis is similar to earlier, the exterior covariant derivative yields extra effects, as was the case for the similar procedure in gravity. The second equation can be formally solved by introducing a Lie algebra valued 3-form $X_a$ such that $\mathrm{D}X_a=0$; an integration constant, so to say. The formal solution
\begin{equation}\label{eq:vector_eom}
	\varepsilon_{abcd}\mathrm{D}\phi^b\wedge e^c\wedge e^d+e_a\wedge F=X_a,\ \mathrm{D}X_a=0,
\end{equation}
can be contracted from the left or right by $\phi^a$. Then subtracting yields the commutator
\begin{equation}
	\varepsilon_{abcd}[\phi^a,\mathrm{D}\phi^b]\wedge e^c\wedge e^d+ e_a\wedge [\phi^a,F]=[\phi^a, X_a].
\end{equation}
Therefore the Yang-Mills equation prototype includes an arbitrary 3-form integration constant in the commutator,
\begin{equation}
	[\phi^a,X_a]+\mathrm{D}(\mathrm{D}\phi^a\wedge e_a)=J.
\end{equation}
In case of Abelian groups $[\phi^a,X_a]=0$, while otherwise this term is nontrivial.

Establishing correspondence between $\mathrm{D}\phi^a\wedge e_a$ and the dual field strength 2-form proceeds in analogy with the previous sections, but in the presence of the extra 3-form $X_a$. For convenience, let $\mathrm{D}\phi^a=u^a$. Investigating the dual density of Eq.~\eqref{eq:vector_eom} starts from
\begin{equation}
	\varepsilon_{abcd} u^b{}_i\varepsilon^{icdk}+\frac{1}{2}F_{cd}\varepsilon_a{}^{cdk}=\frac{1}{3!}X_{aicd}\varepsilon^{icdk}
\end{equation}
and results in
\begin{equation}
	2 u_{ka}= \frac{1}{2}F^{cd} \varepsilon_{cdak}-\frac{1}{3!}X_a{}^{icd}\varepsilon_{icdk}, \label{phieq1}
\end{equation}
so in global form
\begin{equation}
	u_a\wedge e^a=*F+\frac{1}{2}(*X_a)\wedge e^a. \label{phieq2}
\end{equation}
Therefore, the equations of motion of the vector potential $A$ are
\begin{equation}
	[\phi^a,X_a]
	+\frac{1}{2}\mathrm{D}(*X_a\wedge e^a)
	+\mathrm{D}*F=J.
\end{equation}
The theory is equivalent to usual Yang-Mills theory when
\begin{equation}\label{eq:general_vector_eom}
	[\phi^a,X_a]
	+\frac{1}{2}\mathrm{D}(*X_a\wedge e^a)=0.
\end{equation}
A particular solution is $X_a=0\Rightarrow\mathrm{D}X_a=0$, so a proper Yang-Mills limit exists. A solution for the $\phi^a$ should generally exist,
since (\ref{phieq2}) has the same number of equations as unknowns. Looking at this a bit more explicitly, in the geometric phase we can write the components of $u_{ka}$ in (\ref{phieq1}) in some coordinate system as
$u_{[\mu\nu]} = -g_{\alpha[\mu}\nabla_{\nu]}\phi^\alpha$. In the very simplest case of flat space $g_{\mu\nu} = \eta_{\mu\nu}$, $\nabla_\mu = \partial_\mu$, Abelian group $G=U(1)$ and setting $X_a=0$, the solution is simply that $\phi_\mu = \tilde{A}_\mu$ (up to the $U(1)$ ambiguity $\phi_\mu \rightarrow \phi_\mu + \partial_\mu\varphi$), where $\tilde{A}_\mu$ is the electromagnetic gauge field corresponding to the dual field strength. In the generic case the solution for $\phi_\mu = e^a{}_\mu\phi_a$ will be a nonlinear function of the gravitational fields, the gauge fields and the $X^a$-field, but there is no obvious reason why such a solution shouldn't always exist. 
 
In general when $X_a\neq0$ the meaning of the additional terms is not particularly clear. In the case of Abelian $U(1)$ theory, however, there is a simple interpretation in terms of vacuum magnetization and polarization. In usual electromagnetic theory, the current 3-form $J$ splits into the contribution $J^\text{mat}$ from bound electric current inside matter and an external current $J^\text{ext}$ as
\begin{equation}
	J=J^\text{mat}+J^\text{ext}.
\end{equation}
The total current is conserved, $\mathrm{d}J=0$, and it is assumed there is no conversion between internal and external charges. Therefore it is meaningful to introduce a matter excitation $H^\text{mat}$ such that $J^\text{mat}=\mathrm{d}H^\text{mat}$, see Ref.~\cite{Hehl_Obukhov:2003} for details.

This excitation can then be split in terms of magnetization and polarization after proceeding with a 1+3 decomposition. Let spacetime have a local 1+3 foliation, parametrized by a monotonously increasing variable $\sigma$. Therefore topologically\footnote{Let us note in passing that a set of assumptions, like connectedness, orientability, paracompactness, and Hausdorff separability, would be closely related to the existence of a $3+1$ foliation, and further to the existence of a pseudo-Riemannian structure on the manifold. Here we only assume and proceed with the spacetime decomposition to clarify the meaning of the extra terms.} let the differentiable manifold $M=\Sigma\times\mathbb{R}$. The vector field $n$ corresponding to a congruence of observer worldlines is defined by
\begin{equation}
	n\lrcorner\mathrm{d}\sigma=\mathcal{L}_n\sigma=-1.
\end{equation}
Any $p$-form $\omega$ can be split into a component longitudinal to $n$ by
\begin{equation}
	{}^\perp\omega=\mathrm{d}\sigma\wedge\omega_\perp,\quad\omega_\perp=n\lrcorner\omega,
\end{equation}
the remainder being the transverse component,
\begin{equation}
	\underline{\omega}=(\mathds{1}-{}^\perp)\omega=n\lrcorner(\mathrm{d}\sigma\wedge\omega),\quad n\lrcorner\underline{\omega}=0.
\end{equation}
Therefore
\begin{equation}
	\omega={}^\perp\omega+\underline{\omega}=\mathrm{d}\sigma\wedge\omega_\perp+\underline{\omega}.
\end{equation}
Applying this procedure to the internal excitation
\begin{equation}
	H^\text{mat}={}^\perp H^\text{mat}+\underline{H}^\text{mat}=-\mathcal{H}^\text{mat}\wedge\mathrm{d}\sigma+\mathcal{D}^\text{mat}.
\end{equation}
serves as the basis for defining the polarization 2-form $P$ and magnetization 1-form $M$:
\begin{align}
	\mathcal{D}^\text{mat}&=-P,\\
	\mathcal{H}^\text{mat}&=M,
\end{align}
where the minus sign is convention.

In the Abelian case, Eq.~\eqref{eq:general_vector_eom} reduces to
\begin{equation} \label{EMeom}
	\mathrm{d}*F=J-
	\frac{1}{2}\mathrm{d}(*X_a\wedge e^a),
\end{equation}
and we find $\frac{1}{2}\mathrm{d}(*X_a\wedge e^a)$ is precisely of the current form $J=\mathrm{d}H$, with the ``vacuum excitation''
\begin{equation}
	H^\text{vac}=\frac{1}{2}*X_a\wedge e^a. \label{internalexcitation}
\end{equation}
This can then be split into the magnetization and polarization components, as
\begin{eqnarray} \label{Mvac}
M^{\text{vac}} & = & \frac{1}{2}(* X_a(n)\underline{e}^a - e^a(n)\underline{*X}_a)\,, \\
P^{\text{vac}} & = & \frac{1}{2}\underline{e}^a\wedge \underline{*X}_a\,, \label{Pvac}
\end{eqnarray}
where $\underline{e}^a$ is the spatial coframe.
In essence we have found that a suitable reformulation of electromagnetism allows for magnetization and polarization to appear simply as integration constants, similarly to how dark matter is described in the Khronon theory proposed in Ref.~\cite{Zlosnik_et_al:2018}. Defining analogues of magnetization and polarization for non-Abelian Yang-Mills theory is not conventional, however. 

By defining the dressed field strength, $\mathcal{F}=F-*H^{\text{vac}}$, we can rewrite
equation (\ref{EMeom}) as
\begin{equation} \label{dressed}
\mathrm{d}*\mathcal{F} = J\,. 
\end{equation}
By straightforward manipulations, we can show that in the geometric phase the energy-momentum tensor of the generic Yang-Mills theory becomes
\begin{equation}
T^{\mu\nu} = \text{Tr}\left[\frac{1}{2}\left( \mathcal{F}^\mu{}_\alpha \mathcal{F}^{\nu\alpha} 
+ \mathcal{F}^{(\mu}{}_\alpha {F}^{\nu)\alpha}\right) - \frac{1}{4}g^{\mu\nu} \mathcal{F}_{\alpha\beta} {F}^{\alpha\beta}\right]\,. 
\end{equation}
The vacuum excitation modifies the energy-momentum sources in an interesting way.
It may break the conformal invariance of the gauge theory if $T^\mu{}_\mu = -\frac{1}{2}\text{Tr}(*F^{\mu\nu}H^{\text{vac}}_{\mu\nu}) \neq 0$.
If the field strength vanishes, $F_{\mu\nu}=0$, the vacuum may still contain energy due to the $H^{\text{vac}}_{\mu\nu}$. On the other hand, the solution for the gauge field strength $\mathcal{F}_{\mu\nu}=0$ is always available in the absence of material sources $J^\mu=0$, and this solution has zero energy. In the next section \ref{khrononplusiso} we will see that when coupled to Khronon gravity, the $X^a$-field can further generate a ``hypermomentum'' source.

The resulting magnetization and polarization is not completely arbitrary, but is constrained by $\mathrm{D}X_a=0$. It would be attractive to interpret this in terms of ``covariantly closed'' forms, but as in general $\mathrm{D}^2\neq0$, this does not produce a proper cohomology theory. Rather, $\mathrm{D}X_a=0$, and generally $\mathrm{D}\omega=0$ for arbitrary forms $\omega$, could be taken as the generalization of requiring $X_a$ or $\omega$ to be covariantly constant, compare with the exterior covariant derivative $\mathrm{D}$ on some vector bundle $E$ mapping $\mathrm{D}:v\in\mathrm{TM}\mapsto\mathrm{D}_v$, such that for any section $s\in\Gamma(E)$ and vector field $v$ we have $\mathrm{D}s(v)=\mathrm{D}_vs$, among other axioms. Therefore the difficulty of solving $\mathrm{D}X_a=0$, e.g. in terms of differential equations, should be of the same class as finding covariantly constant fields, possibly devolving into (numeric) integration in charts.

The interior product yields the precise relation of $X_a$ to the vacuum excitation 2-form $H^\text{vac}$, thus to the magnetization and polarization. Let $\tetrad_a$ correspond to the vector basis dual to the coframe $e^a$, that is $e^a(\tetrad_b)=\delta^a_b$. Then directly
\begin{equation}
	H^\text{vac}=\frac{1}{2}H_{ij}e^i\wedge e^j=-\frac{1}{2}(\tetrad_a\lrcorner H^\text{vac})\wedge e^a=\frac{1}{2}*X_a\wedge e^a.
\end{equation}
Therefore
\begin{equation}
	X_a=-*(\tetrad_a\lrcorner H^\text{vac}).
\end{equation}
In the regular metric phase, instead of $X_a$, we could consider the Hodge dual $Y_a=*X_a$ as the introduced fundamental quantity. The condition $\mathrm{D}X_a=0$ reads in terms of the excitation as
\begin{equation}
	\mathrm{D}(*(\tetrad_a\lrcorner H^\text{vac}))=0,
\end{equation}
which is rather a co-covariant constancy condition, if a covariant codifferential $\delta_D=*\mathrm{D}*$ were to be introduced. The interpretation of terms in the non-Abelian case remains unclear, however. From the above we see that though the
dressed field strength $\mathcal{F}=F-*H^\text{vac}$ satisfies (\ref{dressed}) (in the premetric context called the first fundamental equation), only the contracted field strength $ \mathcal{F}_a=\tetrad_a\lrcorner F - *(\tetrad_a\lrcorner H^\text{vac})$ would satisfy the adapted 
Bianchi identity (respectively, the second fundamental equation), $\textrm{D}\mathcal{F}_a = -{}^\#T_a\lrcorner F =0$ in the absence of torsion (c.f. Lemma 1 of Ref.~\cite{BeltranJimenez:2020sih}).

\subsection{Pregeometric Yang-Mills theory and Khronon gravity}
\label{khrononplusiso}

The Cartan Khronon theory of gravity is based upon a new approach to the problem of time. Space and time emerge in a spontaneous symmetry breaking which might ultimately be connected to the collapse of the wavefunction.  
At the formal level, the key is the reduction of the coframe to the Cartan Khronon field $\tau^a$ such that $e^a=\mathrm{D}\tau^a$ ~\cite{Zlosnik_et_al:2018} (we will briefly discuss the spinor version of the formulation in Section \ref{sec:unification_questions}). Then, a canonical clock field is built into the structure of the theory, and rather than introducing some $\sigma$ by hand in the decomposition introduced 
above, we can identify $\sigma = i\sqrt{\eta_{ab}\tau^a\tau^b}$. The coupling of the pregeometric Yang-Mills theory to the Cartan Khronon gravity reveals further physical properties of the newly found 3-form $X_a$.

Consider the coupled theory
\begin{equation}
    S = \int \textrm{D}\tau^a\wedge\textrm{D}\tau^b\prescript{+}{}R_{ab} + \int \text{Tr}\left[ \textrm{D}\phi^a\wedge\textrm{D}\tau^b\left( \frac{1}{2} \varepsilon_{abcd} \textrm{D}\phi^c\wedge\textrm{D}\tau^d + \eta_{ab}F\right)\right]\,.
\end{equation}
The equations of motion for the (Cartan) Khronon and the gravitational connection are, respectively,
\begin{eqnarray}
\textrm{D}\left( \prescript{+}{}R^a{}_b\wedge\textrm{D}\tau^b - \theta^a\right) & = & 0\,, \\
\frac{1}{2}\textrm{D}\prescript{+}{}{}\left( \textrm{D}\tau^{[a}\wedge\textrm{D}\tau^{b]}\right)
& = & \tau^{[a}\prescript{+}{}R^{b]}{}_c\wedge\textrm{D}\tau^c - \tau^{[a}\theta^{b]} - \textrm{Tr}\left(\phi^{[a}X^{b]}\right)\,. 
\end{eqnarray}
By integrating the first equation, we obtain the dark matter 3-form $M^a$ such that $\textrm{D}M^a=0$. Using this solution to simplify the second equation, we get
\begin{eqnarray}
\prescript{+}{}R^a{}_b\wedge\textrm{D}\tau^b & = & \theta^a + M^a\,, \label{xeq1} \\
\frac{1}{2}\textrm{D}\prescript{+}{}{}\left( \textrm{D}\tau^{[a}\wedge\textrm{D}\tau^{b]}\right)
& = & \tau^{[a}M^{b]} - \textrm{Tr}\left(\phi^{[a}X^{b]}\right)\,. \label{xeq2}
\end{eqnarray}
In the second equation the LHS is self-dual, and thus must be the RHS. Assuming that this applies to each term separately, we see that whereas the dark matter 3-form satisfies the self-duality condition $\prescript{-}{}(\tau^{[a} M^{b]})=0$ wrt the Khronon $\tau^a$, the vacuum excitation 3-form satisfies the self-duality condition $\textrm{Tr}\prescript{-}{}{}(\phi^{[a}X^{b]})=0$ wrt the iso-Khronon $\phi^a$.  
This justifies the interpretation of the cosmological dark matter as the gravitational analogy to the vacuum magnetization/polarization in the internal gauge theory. 

Let us consider the situation that the Khronon and the iso-Khronon are aligned, i.e. $\phi^a \sim \tau^a$ (obviously, this implies the isotropy of $\phi^a$ in the internal space).
It immediately follows that $\prescript{-}{}(X^{[a}\tau^{b]})=0$.
Then, one can deduce the two consequences of the conservation equation $\mathrm{D} X_a = 0$ (see~\cite{Gallagher:2021tgx}). Firstly, the 3-form $X_a$ is a purely spatial volume form. Secondly, its volume integral is a constant. However, in this case the physical effect of vacuum excitation vanishes, even though the 3-form $X_a$ may exist as a non-trivial 3-form. This is most easily seen in the time gauge $\tau^a=\tau\delta^a_0$, where we may write $X_0 = X \star e_0$ for some scalar $X$, and $X_I =0$, and obtain that $\mathrm{d}X_0=0$. Notationwise, $\star$ is the internal dual, and the capital Latin letters are used for the spatial Lorentz indices. But firstly, we see that the time-like coframe $e^0$ is purely longitudinal $\underline{e}^0=0$, since $e^0 = \mathrm{D}\tau^0 = \mathrm{d}\tau = -\mathrm{d}\sigma$. Secondly, the
only non-vanishing component of $*X_a$ is $*X_0 = * X \epsilon_{IJK}e^I\wedge e^J \wedge e^K/6 =X( *\star e_0)/6 = X e^0$ is also longitudinal. So the $H^{\text{vac}}$ in (\ref{internalexcitation}) vanishes. Thus, the components of $X_a$ that satisfy the self-duality condition with respect to the Cartan Khronon $\tau^a$ do not result in vacuum magnetization or polarization. In particular, if the iso-Khronon $\phi^a$ picks up the time direction preferred by the symmetry-breaking field $\tau$, the effect of $X_a$ is trivialised.

When this is not the case, the theory predicts also novel gravitational effects due to the vacuum excitation.
From (\ref{xeq1}) we see that the Yang-Mills fields contribute to the energy-momentum and thus source gravity as usual. However, there is different kind of contribution in (\ref{xeq2}). Though it appears to be similar to the effect of dark matter, under closer inspection it turns out that this is not the case. Again, it is useful to adapt the 
system into the time gauge $\tau^a=\tau\delta^a_0$. In this gauge, the components of the self-dual curvature reduce to the triad of curvature two-forms $R^I$. The independent components of the anti-self dual curvature are then encoded into the triad of torsion 2-forms $T^I$. In the end, the field equations (\ref{xeq1},\ref{xeq2}) can
be re-expressed in the gauge-fixed form 
\begin{eqnarray}
R_I\wedge e^I = -i\theta^0 -iM^0\,, \\
R^I\wedge\textrm{d}\tau + i\epsilon^I{}_{JK} R^J\wedge e^K & = & -\theta^I\,, \\
T^I\wedge\textrm{d}\tau - i\epsilon^I{}_{JK}T^J\wedge e^K & = & 2\textrm{Tr}\left(\phi^{[I}X^{0]}\right)\,.
\end{eqnarray}
The two first equations above are the energy and the momentum equations, respectively. 
As expected, the dark matter 3-form is associated with effectively pure energy density, and its effective pressure is identically zero.  
The new effect of the excitation 3-form $X^a$ is apparent in the last equation, where it appears as a source of torsion. Thus, the Yang-Mills vacuum excitation can generate nontrivial gravitational ``hypermomentum'' \cite{Hehl:1978cb}. This effect disappears when the 3-form $X^a$ is aligned with the iso-Khronon such that $\phi^a \sim X^a$. The phenomenological implications of the Yang-Mills hypermomentum would be very interesting to explore, but we must leave that for future studies. 

\subsection{Scalar fields}
The scalar field action in a singular metric regime runs into the same problem as  the Yang-Mills action. Similarly, a first order theory can be defined for the field $\phi$ by introducing an auxiliary field $G^{abc}$ with totally antisymmetric indices,
\begin{equation}
	S=\int\bigg(\bigg(\frac{1}{4} G_{abc}G^{abc}+U(\phi)\bigg)\varepsilon_{ijkl}e^i\wedge e^j\wedge e^k\wedge e^l+G_{abc}e^a\wedge e^b\wedge e^c\wedge\mathrm{d}\phi\bigg)
\end{equation}
again producing a 2-step kinetic cycle. Varying by $\phi$ and $G^{abc}$ produces respectively
\begin{align}
	\mathrm{d}(G_{abc}e^a\wedge e^b\wedge e^c)+U'(\phi)\varepsilon_{abcd}e^a\wedge e^b\wedge e^c\wedge e^d&=0,\\
	\frac{1}{4}G^{abc}\varepsilon_{ijkl}e^i\wedge e^j\wedge e^k\wedge e^l+e^a\wedge e^b\wedge e^c\wedge\mathrm{d}\phi&=0,
\end{align}
which is the prototype wave equation and the auxiliary equation. Everything goes by the usual procedure outlined earlier, enforcing
\begin{equation}
	*\mathrm{d}\phi=\frac{1}{3!}\mathrm{d}^k\phi\varepsilon_{kabc}e^a\wedge e^b\wedge e^c,
\end{equation}
and yielding the wave equation
\begin{equation}
	-\mathrm{d}*\mathrm{d}\phi+U'(\phi)\varepsilon_{abcd}e^a\wedge e^b\wedge e^c\wedge e^d=0.
\end{equation}
Therefore all bosonic actions have a pregeometric first order formalism readily available. The interest is then of building a good theory of pregeometry.

\section{\label{sec:unification_questions}Path to Unification}
First order formalism in gravity, that is the Palatini formulation in terms of (co)frames and connections, is well known and has been extensively studied, while the Yang-Mills analogue does not appear to be as popular. It is worth emphasizing how similar these theories appear to the exterior formulation of complex self-dual GR, while the remaining anti-self-dual component is appealing for unification attempts.

Complex GR considers the complexified tensor bundle
\begin{equation}
	T_C=\bigoplus_{r,s}T^r_s(M)\otimes\mathbb{C}
\end{equation}
over a real manifold $M$, see e.g. Ref.~\cite{Giulini:1994} for an overview. The structure group becomes $SO(1,3)_\mathbb{C}\cong SO(4)_\mathbb{C}$, while the fields become complex-valued, i.e. sections of a complex bundle. In addition to investigating Hodge dualization and its eigenforms, we can define a dualization operation $\star$ in the complexified Lie algebra $\mathfrak{so}(1,3)_\mathbb{C}$ such that
\begin{equation}
	\star\omega_{ab}=\frac{1}{2}\epsilon_{ab}{}^{cd}\sigma_{cd},
\end{equation}
decomposing the Lie algebra into self-dual anti-self-dual components
\begin{equation}
	\mathfrak{so}(1,3)_\mathbb{C}=\mathfrak{so}(1,3)_\mathbb{C}^{+}\oplus\mathfrak{so}(1,3)_\mathbb{C}^{-},
\end{equation}
such that
\begin{equation}
	\mathfrak{so}(1,3)_\mathbb{C}^{\pm}=\{\omega\in\mathfrak{so}(1,3)\vert\star\omega=\pm i\omega\}.
\end{equation}
The corresponding projector is
\begin{equation}
	P^{\pm}=\frac{1}{2}(\mathds{1}\mp i\star).
\end{equation}
Note $\mathfrak{so}(1,3)_\mathbb{C}^{(\pm)}$ are simple Lie algebras, while $\mathfrak{so}(1,3)_\mathbb{C}$ is semi-simple. Significantly, the Palatini action
\begin{equation}
	S_\mathbb{C}=\frac{1}{2\kappa}\int *(e^a\wedge e^b)\wedge R_{ab}
\end{equation}
decomposes into
\begin{equation}
	S_\mathbb{C}=S_\mathbb{C}^+ + S_\mathbb{C}^-=\frac{i}{2\kappa}\int e^a\wedge e^b\wedge{}^+R_{ab} - \frac{i}{2\kappa}\int e^a\wedge e^b{}\wedge^-R_{ab}
\end{equation}
and it suffices to consider only one of the actions in the decomposition, as the stationary points of $S_\mathbb{C}$ and $S_\mathbb{C}^\pm$ lie over the same coframe fields. Working with the self-dual action is also the basis for Ashtekar's variables~\cite{Ashtekar:1986, Ashtekar:1987}, establishing phase-space correspondence with $SU(2)$ Yang-Mills theory.

The natural continuation is with the cosmological constant $\Lambda$, which makes the similarity with Yang-Mills theory plain. The complete first order $\Lambda$-Einstein-Yang-Mills theory action is then,
\begin{equation}
	\begin{aligned}
		S&=
		\begin{aligned}[t]\frac{1}{2}\int\mathrm{Tr}_G\bigg[\kappa^{-1}&\bigg(-2\Lambda\epsilon_{abcd}e^a\wedge e^b\wedge e^c\wedge e^d + e^a\wedge e^b\wedge i{}^+R_{ab}\bigg)+\\
		+&\bigg(\frac{1}{2}\,\epsilon_{abcd}u^a\wedge e^b\wedge u^c\wedge e^d + u^a\wedge e^b\wedge \eta_{ab}F\bigg)\bigg]=
		\end{aligned}\\
		&=\frac{1}{2}\int\mathrm{Tr}_G\bigg[(i\kappa^{-1} e^a\wedge e^b + u^a\wedge e^b) \wedge \bigg({}^+R_{ab} + \eta_{ab}F + \epsilon_{abcd}\left(2i\Lambda e^c\wedge e^d  +\frac{1}{2} u^c\wedge e^d\right)\bigg)\bigg]
	\end{aligned}
\end{equation}
although the use of purely self-dual surface elements ${}^+(e^a\wedge e^b)$ in the GR action is conceivable as well. The Yang-Mills action can include a dimensionless constant, which is here set to unity. The many terms in the action can be grouped in various ways. For instance, in principle $\eta_{ab}$ could be put together with the vector potential $A\to\eta_{ab}A$, as to correspond to trace components in a connection-like 1-form
\begin{equation}
	\tilde\omega_{ab}={}^+\omega_{ab}+\eta_{ab}A,
\end{equation}
and furthermore
\begin{equation}
	\tilde{R}_{ab}=\mathrm{d}({}^+\omega_{ab}+\eta_{ab}A)+({}^+\omega_{ac}+\eta_{ac}A)\wedge({}^+\omega^c{}_b + \delta^c_b A)={}^+R_{ab} + \eta_{ab}F.
\end{equation}
When restricting to the case $G=U(1)$, the gravitoelectromagnetic unification is achieved neatly, since $\tilde{R}_{ab}$ is precisely the (self-dual projection of the) Weyl extension of the Lorentz curvature. However, when considering the more complete unification along the lines e.g. $SO(3,13) \rightarrow SO(1,3)\times SO(10)$~\cite{Percacci:2009,Krasnov:2017epi}, which appears quite natural and attractive in this context~\cite{Wilczek:1998}, the generalization of the Weyl 1-form $A$ into the adjoint of a non-Abelian algebra such as the $SO(10)$ forces to rethink the most conventional GraviGUT schemes. The division into the surface element $e^a\wedge e^b$, resp. $u^a\wedge e^b$, and gauge curvature ${}^+R+F$ is clear, however. If $\Lambda=\frac{1}{4}\kappa^{-1}$, then the replacements
\begin{equation}
	i\kappa^{-1}e^a\wedge e^b\to u^a\wedge e^b,\quad {}^+R_{ab}\to\eta_{ab}F
\end{equation}
would transform the gravitational action to a Yang-Mills one. It would be noteworthy to formulate this as a rigorous gauge principle, similar to how fermionic fields couple to gauge bosons, but this doesn't appear simple or unambiguous. 

The value of the cosmological constant is a problem, or alternatively a hint of the precise structure of the underlying theory. Generally it is expected to be of QFT origin, but the calculated value so far is sharply disconnected from measurement, see e.g. Ref.~\cite{Weinberg:1989} for an overview. However, if it is to believed that gravity forms a unified theory with the Standard Model (or a suitable extension), then this analysis supports a fundamental origin for the cosmological constant, possibly related to symmetry breaking.

The gauge group trace really only applies to the Yang-Mills term, but it is formally possible to introduce traceless Lorentz generators in the vector representation
\begin{equation}
	(J^{ab})^{cd}=\frac{i}{2}(\eta^{ac}\eta^{bd} - \eta^{ad}\eta^{bc}),
\end{equation}
so that the Lorentz trace of the product
\begin{equation}
	\mathrm{Tr}(J_{ab}J^{uv})=(J_{ab})^i{}_j(J^{uv})^j{}_i=
	\delta_a^{[u}\delta_b^{v]},
\end{equation}
and the total action involves both traces,
\begin{equation}
	S=\frac{1}{2}\int\mathrm{Tr}\bigg[(e^a\wedge e^b)(i\kappa^{-1}J_{ab}+G_{ab})\wedge \bigg({}^+R_{ij}J^{ij} + F + (e^c\wedge e^d)\epsilon_{ijcd}\left(2i\Lambda J^{ij}+\frac{1}{4}G^{ij}\right)\bigg)\bigg],
\end{equation}
implying that the proper path forward would be through some Lie algebra scheme, particularly when separating the (here) $\Lambda=\frac{1}{8}\kappa^{-1}$ component. In this case, introducing an infinitesimal-like transformation
\begin{equation}
	k_{ij}{}^{ab}=i\kappa^{-1}\delta_i^a\delta_j^b+\frac{1}{4}G_{ij}\eta^{ab}+\frac{1}{4}\eta_{ij}G^{ab},
\end{equation}
yields the action
\begin{equation} \label{surface_action}
	S=\frac{1}{2}\int\mathrm{Tr}\bigg[(e^a\wedge e^b)k_{ab}{}^{ij}\wedge \bigg({}^+R_{ij} + \eta_{ij}F + k_{ij}{}^{uv}(e^c\wedge e^d)\epsilon_{cduv}\bigg)\bigg],
\end{equation}
and the interpretation is that $G^{ab}$ is an infinitesimal-like surface excitation, or the surface element $e^a\wedge e^b$ is transformed, such that the invariant part corresponds to gravity and the change to Yang-Mills theory.

It is ambiguous which is the best interpretation for the auxiliary field, thus also leaving ambiguity in how exactly a unified formulation should arise. In the various formulations in this paper, the new field can be construed as either a 0-form or 1-form field and potentially physically meaningful or not; as a Yang-Mills charged coframe; as a linear transformation between Lorentz and Yang-Mills algebra; as Lagrange multiplier-like fields; as dynamical components of the dual field strength; as substructure of the electromagnetic excitation; as an additional set of vectors to the Khronon of Ref.~\cite{Zlosnik_et_al:2018}; as an infinitesimal Yang-Mills charged surface excitation. At the very least, in a Gravity - Standard Model unification, it can be assumed that the first order formulation represents the symmetry broken phase.

In another approach, the 3 leftover anti-self-dual generators could be mapped to Yang-Mills generators \cite{Gallagher:2021tgx}. For instance, it is possible to introduce a Yang-Mills charged ``mixing matrix'' $z_{ab}$, in particular the complex setting of electromagnetic $U(1)$ theory already appears in the complexified gravity formulation. So it could be defined
\begin{equation}
	A=z^{ab}{}^-\omega_{ab},\ F=\mathrm{d}(z^{ab}{}^-\omega_{ab}).
\end{equation}
In so simple a formulation, this is not a proper unified theory as commonly understood~\cite{Percacci:2009}, i.e. roughly where the vacuum expectation value of a given order parameter selects commuting subgroups of a larger gauge group. Alternatively, if instead the target was $SU(2)$ of the weak interaction, this could be adapted to some alternative of graviweak unification; a proper graviweak formalism is e.g. discussed in~\cite{Nesti_Percacci:2008}, considering $SO(4)_\mathbb{C}$ and the soldering form as an order parameter.

The theory of gravity is quite rich in similar theories. In addition to Palatini formalism, Plebanski formalism~\cite{Plebanski:1977}
\begin{equation} \label{plebanski}
	S_\text{Plebanski}=\int\bigg(B^{ab}\wedge R_{ab} - \frac{1}{2}\phi_{abcd}B^{ab}\wedge B^{cd}\bigg)
\end{equation}
has to be noted as well. In comparison, the surface element is replaced with a single 2-form $e^a\wedge e^b\to B^{ab}$ and fixed by essentially a Lagrange multiplier
\begin{equation}
	\phi_{abcd}=\phi_{cdab}=-\phi_{bacd},
\end{equation}
which enforces that on-shell $B^{ab}$ agrees with ${}^+(e^a\wedge e^b)$; see also Refs.~\cite{Krasnov:2018, Freidel_Speziale:2012} and the references therein. Plebanski theory was likewise utilized in unification~\cite{Smolin:2018}, where the embedding into a larger gauge group and the addition of an extra term produces $GR$ with a Yang-Mills action. In some spin foam approaches to quantum gravity, one  considers only the first term in the Plebanski action (\ref{plebanski}), and may then at a suitable point impose the so called simplicity constraint that reduces the two-form to the exterior product of the tetrad one-forms~\cite{Baez:1997zt}. While this establishes an appealing connection between gravity and topological QFT without local degrees of freedom, the tetrad is a quite complicated, mixed-index 16-component object that is completely alien to standard Yang-Mills theory. The insight of Ref.~\cite{Zlosnik_et_al:2018} that the tetrad should rather emerge as a covariant derivative of a (Lorentz-charged) Higgs-like scalar is not new, but goes back, via Akama~\cite{Akama:1978} and others to the original generalization of F. Klein's geometrical framework to describe symmetries of physics by {\'E}. Cartan~\cite{cartan1986manifolds}. Since we can reduce the Lorentz-charged scalar to a Dirac bispinor, the possibility arises that the metric structure could be reduced to a property of matter fields. Further, the dynamics for this metric structure could arise solely from the quantum fluctuations of matter fields, as famously shown by Sakharov~\cite{Sakharov:1967pk}. Perhaps all the fundamental interactions emerge in a similar fashion from a purely material origin?

To recapitulate the above-described ``series of further and yet further simplicity constraints'' that connects the topological BF theory to what we may call the Dirac-Cartan Khronon gravity:
\begin{equation}
B^{ab} \rightarrow  e^a\wedge e^b \rightarrow \mathrm{D}\tau^a\wedge \mathrm{D}\tau^b \rightarrow \mathrm{D}(\bar{\psi}\gamma^a\psi)\wedge \mathrm{D}(\bar{\psi}\gamma^b\psi)\,. \label{chain}  
\end{equation}    
Let us just in passing mention that the gravity theory~\cite{Zlosnik_et_al:2018} has the remarkable formulation, which is both quartic in the primordial spinor and quartic in the gauge-covariant derivative\footnote{In our convention, the operator $\mathrm{D}$ acts only to the right. In an alternative convention, e.g. $\mathrm{D}(\bar{\psi}\gamma^a\psi)$ in (\ref{chain}) would read $\bar{\psi}\gamma^a\overset{\leftrightarrow}{\mathrm{D}}\psi$.}
\begin{equation} \label{primordial}
S = i\int (\bar{\psi}\gamma_a\psi) \mathrm{D} \prescript{+}{}{\mathrm{D}} \prescript{+}{}{\mathrm{D}} \mathrm{D} (\bar{\psi}\gamma^a\psi)\,.     
\end{equation}
A partial integration and a couple of steps back in the chain (\ref{chain}) brings this into the familiar self-dual Palatini form. Throughout this article, we've had in mind that the more fundamental formulation of gravity should be considered in terms of the primordial $\psi$ rather than $e^a$. As we saw in Section \ref{khrononplusiso}, coupling to Khronon gravity can reveal further properties of the pregeometric Yang-Mills theory.

Of course, any unification attempt with gravity and particle physics is constrained by various theorems, including the commonly cited Coleman-Mandula theorem~\cite{Coleman_Mandula:1967}, which implies that the symmetry group of the underlying QFT can only be the direct product of the Poincar{\'e} and an internal symmetry group. Although the assumptions of the Coleman-Mandula theorem appear natural, the common implication is not unavoidable. Ref.~\cite{Smolin:2018} avoids it as the Coleman-Mandula theorem requires the $S$-matrix symmetries include global Poincare invariance, while the proposal held no global symmetries; likewise in the broken phase of Ref.~\cite{Nesti_Percacci:2008}, the residual symmetry is precisely the required global Lorentz and local internal symmetry --- the Coleman-Mandula theorem requires the existence of a Minkowski metric, while in the pregeometric regime with a vanishing soldering form, there is no explicit metric on the manifold. This also agrees with our discussion, implying that first order theory arises naturally in the broken phase.

\section{\label{sec:summary}Conclusion}
The appearance of Hodge dualization in the actions describing matter and gravity can be avoided by using the first order formalism. For gravity, this proceeds from Palatini to a self-dual formulation. Spinor fields themselves require no inverse metric when explicitly working with $\mathfrak{so}(1,3)$ indices. For the remaining bosonic scalar and Yang-Mills actions, the polynomial first order formulation then goes through a two-step process, such that the usual wave or Yang-Mills equations appear on-shell, while reference to Hodge dualization is in effect replaced with $\mathfrak{so}(1,3)$-dualization, with no necessity of the inverse metric. The results obtained are consistent with those of the usual theories at a classical level. Consistency at the quantum level has been earlier investigated in usual index-notation approaches, and, depending on the precise formulation, is either immediately applicable or expected to hold barring the gravitational sector. Notably, we connect the fundamental axiom of charge conservation (and thus the appearance of the electromagnetic excitation) in the premetric programme to a fundamental field in the pregeometric programme.

Generalizing previous Cartan-geometric results for dark matter and gravity, first order Yang-Mills formalism admits the description where a single Lorentz vector is introduced, rather than a 1-form. This applies to both Abelian and non-Abelian theories, requires introducing the least amount of extra degrees of freedom and produces interesting effects. In the  $U(1)$ theory of electromagnetism, the new terms can be interpreted as vacuum magnetization and polarization, while the generalization of such an interpretation to non-Abelian theory is not conventional. It is unlikely that more minimal schemes of this method exist, at least in 4 dimensions and excepting possible internal changes to the vector, as the method depends on coupling to Levi-Civita symbols via a Lorentz index. At the very least, the vector approach has a consistent phase with usual Yang-Mills theory, and deserves further investigation, e.g. in Cartan geometry.

Comparing with first order formalism in the theory of gravity, there appears a strong case for some kind of dual or unified description of gravitation with gauge theories in complexified theory. However, the precise path remains yet ambiguous, not in small part due to the many possible interpretations of the auxiliary field. It is attractive to interpret it as some Yang-Mills charged coframe, but it could likewise be a transformation between Lie algebras or some excitation. Likewise in this duality, the value of the cosmological constant is curious. It remains to be seen whether this is coincidental or insightful, and what might resolve the problems.

In the context of the results obtained, a natural direction appears to investigate Lie algebras and symmetry breaking, and perhaps dimension-dependence of this formulation. By the work of Ashtekar~\cite{Ashtekar:1986, Ashtekar:1987}, we know that the Hamiltonian form of self-dual GR is closely related to Yang-Mills theory. Likewise we see that the Lagrangian of self-dual GR with the cosmological constant is very similar to that of first first-order Yang-Mills theory. Starting in a complexified first order theory seems promising, with the usual theories possibly only appearing in the end after symmetry breaking and applying suitable reality conditions.

Let us wrap up the article. By construction, QFT is an
effective framework that should robustly approximate physics up to a given energy scale \cite{Weinberg:1996}. A first order reformulation of the Standard Model could be a natural step towards a possibly more fundamental theory. Further, there are rather compelling arguments, beginning from the elementary, classical reasoning that is the basis of the premetric program, and extending to today's cutting-edge speculations about the nature of quantum gravity, that the metric tensor is an emergent field that may even vanish in its ground state. In this cross-lighting, it may seem surprising that a more systematic investigation of pregeometric first order Yang-Mills theory has not yet been undertaken in the Literature.

Our basic finding is that the field excitation tensor $H$ effectively becomes a fundamental field of the gauge theory, on the same footing as the connection $A$ that gives rise to the field strength tensor $F=\mathrm{D}A$. We considered several formulations of this principle, suggesting several new directions to pursue, but they can be all classified according to which kind of field is considered to be the variational degree of freedom.
\begin{itemize}
\item The standard formulation imposes $H=* F$ without dynamical variation.
In the premetric language, this is the axiom of constitutive law.
\item The coframe variation considers $u^a$, and results in
$H \approx u^a\wedge e_a$. Whilst perhaps uneconomical, such theories
suggest interesting connections to bimetric gravity on one side, and to
geometric formulations of QCD on the other.
\item The group element $G_{ab}$ as a variational degree of freedom results in
$H \approx G_{ab}e^a \wedge e^b$. This approach allows the interpretation of the
unified theory (\ref{surface_action}) as a surface excitation of a topological action.
\item The vector substructure $\phi^a$ results in a relation
$H + \mathcal{X} \approx \mathrm{D} \phi^a\wedge e_a$, where $\mathcal{X}$ is
an extra 2-form, that (at least in the electromagnetic case) allows the
interpretation in terms of vacuum magnetization and polarization, as well as its surprising analogy with the cosmological dark matter.
\item The spinor substructure $\psi$ is an alternative to the
vector substructure, based on that $\phi^a \approx \bar{\psi}\gamma^a\psi$.
This is the approach we intend to study at more depth in the future.
\end{itemize}
The unification of the primordial spinor gravity (\ref{primordial}) and the pregeometric Yang-Mills theory reduced to a spinor substructure might be a step towards the lower-level QFT that we have been seeking. In a complementary approach, ascending from the first principles towards the higher level of a dynamical QFT, progress is being made indeed (rather than e.g. qubits) in terms of fermions~\cite{Perinotti:2017neo}. Each brick bridging the gap between these levels is paving the way for a new paradigm.

\begin{acknowledgments}
The authors thank Tom Z\l{}o\'snik for helpful discussions. This work was supported by the Estonian Research Council grants PRG356 ``Gauge Gravity", MOBTT86 and by the EU through the European Regional Development Fund CoE program TK133 ``The Dark Side of the Universe".
\end{acknowledgments}

\bibliography{yangmills_references}

\end{document}